\documentclass[twoside]{article}

\usepackage{blindtext} 

\usepackage[sc]{mathpazo} 
\usepackage[T1]{fontenc} 
\usepackage{microtype} 

\usepackage[english]{babel} 

\usepackage[hmarginratio=1:1,top=32mm,columnsep=20pt]{geometry} 
\usepackage[hang, small,labelfont=bf,up,textfont=it,up]{caption} 
\usepackage{booktabs} 

\usepackage{lettrine} 

\usepackage{enumitem} 
\setlist[itemize]{noitemsep} 

\usepackage{abstract} 

\usepackage{titlesec} 
\renewcommand\thesection{\Roman{section}} 
\renewcommand\thesubsection{\roman{subsection}} 
\titleformat{\section}[block]{\large\scshape\centering}{\thesection.}{1em}{} 
\titleformat{\subsection}[block]{\large}{\thesubsection.}{1em}{} 

\usepackage{fancyhdr} 
\usepackage{titling} 

\usepackage{hyperref} 
\usepackage{authblk}

\usepackage{slashed}
\usepackage{dsfont}
\usepackage{amsfonts}
\usepackage{amsmath}

\pretitle{\begin{center}\Huge\bfseries} 
\posttitle{\end{center}} 
\title{Scalar Quantum Electrodynamics via Duffin-Kemmer-Petiau Gauge Theory in the Heisenberg Picture:Vacuum Polarization} 
\author[1]{J. Beltran\thanks{jhosep@ift.unesp.br}}
\author[1]{N. T. Maia\thanks{nmaia@ift.unesp.br}}
\author[1]{B. M. Pimentel\thanks{pimentel@ift.unesp.br}}
\affil[1]{S\~{a}o Paulo State University (UNESP), Institute for Theoretical Physics (IFT), R. Dr. Bento Teobaldo Ferraz 271 CEP 01140-070, S\~{a}o Paulo, SP}
\date{\today} 
\renewcommand{\maketitlehookd}{%
\begin{abstract}
\noindent Scalar Quantum Electrodynamics is investigated in the Heisenberg picture via the Duffin-Kemmer-Petiau gauge theory. On this framework, a perturbative method is used to compute the vacuum polarization tensor and its corresponding induced current for the case of a charged scalar field in the presence of an external electromagnetic field. Charge renormalization is brought into discussion for the interpretation of the results for the vacuum polarization. 
\end{abstract}
}

\begin{document}

\maketitle


\section{Introduction}
\label{intro}
There are few approaches to solve Quantum Field Theory in the Heisenberg picture. One of the ways to do this is through the methodology developed by Nakanishi \cite{Nakanishi} where a representation of the operator solution is obtained by constructing the set of all Wightman functions for the fundamental fields. Another framework was developed by K\"{a}ll\'{e}n \cite{Kallen1,Kallen2} which has been applied to the Thirring Model \cite{Lunardi} and Quantum Electrodynamics in three dimensions \cite{Lunardi2}. In this work we will present the methodology of K\"{a}ll\'{e}n and apply it to Scalar Quantum Electrodynamics (SQED) via Duffin-Kemmer-Petiau (DKP) gauge theory.

Despite the most popular approach to confront SQED, which makes use of Klein-Gordon-Fock (KGF) equation, we shall work with the scalar Duffin-Kemmer-Petiau (SDKP) first order differential equation which has the immediate advantage of presenting similitude with Dirac equation. Firstly introduced by Petiau \cite{Petiau}, the DKP matrix algebra was later shown to be decomposed into irreducible representations of 5, 10 and 1 degrees \cite{Geheniau}. Independently, Kemmer wrote Proca equation as a set of coupled first order equations and addressed the spin-0 case \cite{Kemmer0}. Duffin relied on Kemmer's work to write the equations set in a first order matrix formulation endowed by 3 of the 4 commutation relations present in the later DKP algebra \cite{Duffin}. With this result, Kemmer was able to complete the formalism and present the full theory of a relativistic wave equation for bosons of spin 0 and 1 \cite{Kemmer}. The historical development of DKP theory with accurate references can be found in \cite{Nieto,Tati}.

The equivalence between {DKP and KGF theories} for the free case is known, furthermore it was shown that both theories are equivalent in the classical level for the cases of minimal interaction with electromagnetic \cite{CassElDKP} and gravitational fields \cite{Okubo,ClasGrav DKP}. Strict proofs of equivalence between both theories were also given for cases of interaction of quantized scalar field with classical and quantized electromagnetic, Yang-Mills and external gravitational fields \cite{FormalDKP-KG}. Nevertheless, the algebraic structure of the DKP formalism features a variety of couplings that cannot be expressed in KGF and Proca theories \cite{Fisch,Clark}. Studies on spin-0 and spin-1 DKP fields have been done in pure Riemann and Riemann-Cartan spacetimes \cite{r1,j1,r2,r3,a1}. Out of many applications, DKP theory has been used to explore SQED \cite{Umezawa,Akhiezer,Kinoshita1,Kinoshita2,Tomazelli} in Interaction picture and also QCD \cite{Gribov}, as well as applied to covariant Hamiltonian dynamics \cite{Kana} and to the study of spin-1 particles in the Abelian monopole field \cite{Red}.

We start reviewing the free DKP quantum fields in Sect. \ref{FDKP}. In Sect. \ref{FEF} we present the most relevant properties for the quantized electromagnetic field for the aim of this article. The coupling of DKP and electromagnetic fields and the solution of the coupled differential equations by K\"{a}ll\'en's methodology are presented in Sect. \ref{Kall}. In Sect. \ref{VAC} we derive the vacuum polarization tensor and the consequent induced current and discuss charge renormalization. In last section we present the final remarks and prospects.
\section{Duffin-Kemmer-Petiau Gauge Field Theory}\label{FDKP}
We begin by reviewing the DKP gauge theory for charged spin-0 particles. The most used method for studying SQED is based on the Lorentz invariant KGF equation. However, we shall use another approach based on the first order DKP equation
\begin{equation}\label{eqpsi}
(i\beta^\mu\partial_\mu-m)\psi(x)=0,
\end{equation}
where the metric signature is $g^{\mu\nu}=diag(1,-1,-1,-1)$ and the $\beta^\mu$ are square matrices which obey the following algebra
\begin{equation}\label{algebrabeta}
\beta^\mu\beta^\nu\beta^\rho+\beta^\rho\beta^\nu\beta^\mu=\beta^\mu{g}^{\nu\rho}+\beta^\rho{g}^{\mu\nu}.
\end{equation}

The algebra (\ref{algebrabeta}) has three irreducible representations \cite{Petiau,Duffin,Kemmer} of order 1, 5 and 10. For this work we will use the representation of order 5 which represents particles with 0-spin. Therefore, the DKP field $\psi(x)$ is a $5$-component vector with the following conjugation
\begin{equation}\label{conjugadapsi}
\bar{\psi}(x)=\psi^\dag(x)\eta^0,\quad \eta^0=2(\beta^0)^2-1.
\end{equation}

The conjugate field $\bar{\psi}(x)$ obeys the equation
\begin{equation}\label{eqpsibar}
\overline{\psi}(x)(i\beta^\mu\overleftarrow{\partial_\mu}+m)=0.
\end{equation}

The two equations (\ref{eqpsi}) and (\ref{eqpsibar}) can be obtained by the symmetrized Lagrangian density
\begin{equation}\label{LagDkp}
\mathcal{L}_{\text{DKP}}=\frac{i}{2}\overline{\psi}(x)\beta^\mu\overleftrightarrow{\partial_\mu}\psi(x)-m\overline{\psi}(x)\psi(x).
\end{equation}

Solving the classical equation of motion, we obtain for the DKP fields the classical solutions
\begin{equation}\label{DKP geral solution 3}\begin{aligned}
\psi(x)
&=\int\frac{d^3p}{(2\pi)^\frac{3}{2}}a(\mathbf{p})u^-(\mathbf{p})e^{-ipx}+\int\frac{d^3p}{(2\pi)^\frac{3}{2}}b^*(\mathbf{p})u^+(\mathbf{p})e^{ipx},
\end{aligned}\end{equation}
\begin{equation}\label{DKP geral solution 3 conj}\begin{aligned}
\overline{\psi}(x)
&=\int\frac{d^3p}{(2\pi)^\frac{3}{2}}a^*(\mathbf{p})\overline{u^{-}}(\mathbf{p})e^{ipx}+\int\frac{d^3p}{(2\pi)^\frac{3}{2}}b(\mathbf{p})\overline{u^{+}}(\mathbf{p})e^{-ipx},\\
\end{aligned}\end{equation}
where $u^-(\mathbf{p})$ and $u^+(\mathbf{p})$ are 5-component vectors normalized as
\begin{equation}\label{normu}
\overline{u^\pm}\beta^0u^\pm=\mp1.
\end{equation}

The classical 4-current associated with the charged particles can be obtained through the U(1) transformation $\psi'(x)=e^{ie\lambda}\psi(x)$:
\begin{equation}\label{4corriente}\begin{aligned}
j^\mu(x)
&={e}\overline{\psi}(x)\beta^{\mu}\psi(x),\\
\end{aligned}\end{equation}
where $e$ is the coupling constant.

Using the classical solutions (\ref{DKP geral solution 3}) and (\ref{DKP geral solution 3 conj}), we quantize the DKP fields taking the coefficients $a^*(\mathbf{p})$ and $a(\mathbf{p})$ as creation and annihilation operators with the commutation rule
\begin{equation}\label{comopcreani}\begin{cases}
[\hat{a}(\mathbf{p}),\hat{a}^\dag(\mathbf{p}')]=\delta(\mathbf{p}-\mathbf{p}'),\\
[\hat{b}(\mathbf{p}),\hat{b}^\dag(\mathbf{p}')]=\delta(\mathbf{p}-\mathbf{p}').\\
\end{cases}\end{equation}

The quantized current operator can be constructed by symmetrization of (\ref{4corriente}):
\begin{equation}\label{4corrienteq}\begin{aligned}
\hat{j}^\mu(x)&\equiv\frac{e}{2}\{\overline{\psi}_a(x)\beta^{\mu}_{ab},\psi_b(x)\}\\
&=\frac{e}{2}\beta^{\mu}_{ab}\{\overline{\psi}_a(x)\psi_b(x)+\psi_b(x)\overline{\psi}_a(x)\}.\\
\end{aligned}\end{equation}
The symmetrization process fixes the vacuum expectation value of the current
\begin{equation}\label{4corrienteq2}\begin{aligned}
\langle0|j^\mu(x)|0\rangle
&=0,\\
\end{aligned}\end{equation}
which means that, in this case, the symmetrization of the current is equivalent to the normal order.

From the commutation rules (\ref{comopcreani}), we can infer that the commutation of DKP fields is a c-number
\begin{equation}\label{Com final com S}
[\psi_a(x),\overline{\psi}_b(y)]=\frac{1}{i}S_{ab}(x-y),
\end{equation}
where, using the Feynman analog notation $\slashed{V}=V_{\mu}\beta^{\mu}$, the matrix element $S_{ab}(z)$ is
\begin{equation}\label{S}\begin{aligned}
S_{ab}(z)
&\equiv\frac{1}{m}[i\slashed{\partial}(i\slashed{\partial}+m)]_{ab}D_m(z).\\
\end{aligned}\end{equation}

$D_m(x)$ is the Pauli-Jordan causal distribution for a particle of mass $m$
\begin{equation}\label{JPD}\begin{aligned}
D_m(x)
&=\frac{i}{(2\pi)^3}\int\frac{d^3p}{2p^0}\left(e^{-ip_\alpha x^\alpha}-e^{ip_\alpha x^\alpha}\right),\quad p^{0}=E_\mathbf{p}, \quad p^2=m.
\end{aligned}\end{equation}

Another useful operation with DKP fields is their anti-commutation. The latter is not a c-number, but the vacuum expectation value
\begin{equation}\label{Antcom4}\begin{aligned}
\langle0|&\{\bar{\psi}^{(0)}_l(y),\psi^{(0)}_b(x)\}|0\rangle=S^{(1)}_{bl}(x-y),\\
\end{aligned}\end{equation}
where
\begin{equation}\label{S1}
S^{(1)}_{bl}(z)\equiv
\int\frac{d^4p}{(2\pi)^{3}}\delta(p^2-m^2)\frac{1}{m}[\slashed{p}(\slashed{p}+m)]_{bl}e^{-ip(z)}.
\end{equation}

An important point is the solution of the inhomogeneous DKP equation
\begin{equation}\label{equanhomogenea}
\left(i\slashed{\partial}-m\right)\psi(x)=f(x).
\end{equation}

Making use of the algebra (\ref{algebrabeta}) we can compute the retarded and advanced Green functions $G_{R,A}$
\begin{equation}\label{RetG}\begin{aligned}
G_R(x-y)
&=\frac{(2\pi)^{-4}}{m}(i\slashed{\partial}^x)(i\slashed{\partial}^x+m)\int{}d^{4}k[\frac{e^{-ik(x-y)}}{k^2-m^2+ik_00}]-\frac{\delta(x-y)}{m},\\
\end{aligned}\end{equation}
\begin{equation}\label{AvG}\begin{aligned}
G_A(x-y)
&=\frac{(2\pi)^{-4}}{m}(i\slashed{\partial}^x)(i\slashed{\partial}^x+m)\int{}d^{4}k[\frac{e^{-ik(x-y)}}{k^2-m^2-ik_00}]-\frac{\delta(x-y)}{m}.\\
\end{aligned}\end{equation}

The solution of (\ref{equanhomogenea}) can be written as
\begin{equation}\label{solgreenge}\begin{aligned}
\psi(x)=\psi^{(0)}(x)+\int{d^4y}G_R(x-y)f(y),
\end{aligned}\end{equation}
where
\begin{equation}\label{SGequanhomogenea}
(i\slashed{\partial}-m)G_R(x-y)=\delta(x-y)
\end{equation}
and $\psi^{(0)}(x)$ is the solution of the homogeneous DKP equation.

Similarly, for the conjugate DKP equation
\begin{equation}\label{eq mov conpsi2}
\overline{\psi}(x)(i\overleftarrow{\slashed{\partial}}+m)=g(x),
\end{equation}
we get the solution
\begin{equation}\label{solgreengebar}\begin{aligned}
\bar{\psi}(x)
&=\bar{\psi}^{(0)}(x)-\int{d^4y}g(y)G_A(y-x),\\
\end{aligned}\end{equation}
where
\begin{equation}\label{eq mov conpsi2G}
G_A(y-x)(i\overleftarrow{\slashed{\partial}}^{x}+m)=-\delta(y-x).
\end{equation}

\section{Electromagnetic Field}\label{FEF}

For the free electromagnetic field we use the following Lagrangian density
\begin{equation}\label{ElectromagneticL}
\mathcal{L}=-\frac{1}{4}F_{\mu\nu}F^{\mu\nu}-\frac{1}{2}\partial_{\gamma}A^{\gamma}\partial_{\varepsilon}A^{\varepsilon},
\end{equation}
and from this we obtain the equations of motion for the 4-vector potential $A^{\mu}$ are
\begin{equation}\label{EMEL2}\begin{aligned}
\Box{A}^{\mu}&=0.\\
\end{aligned}\end{equation}

As we can see in (\ref{EMEL2}), we have the Klein-Gordon-Fock equation for each component of $A^{\mu}$, then the solution for each component is known. For consistency with the tensorial nature of the commutation of fields, we modify the first term $A^{0}$
\begin{equation}\label{cuantizacion de campo electro componente 0}\begin{aligned}
A^0(x)
&=(2\pi)^{-3/2}\int\frac{{d^3}k}{\sqrt{2E_{\mathbf{k}}}}({\widehat{a}}^0(\mathbf{k})e^{-ikx}-{\widehat{a}}^{0\dag}(\mathbf{k})e^{ikx}),
\end{aligned}\end{equation}
\begin{equation}\label{cuantizacion de campo electro componentes vectoriales}\begin{aligned}
A^i(x)
&=(2\pi)^{-3/2}\int\frac{{d^3}k}{\sqrt{2E_{\mathbf{k}}}}({\widehat{a}}^i(\mathbf{k})e^{-ikx}+{\widehat{a}}^{i\dag}(\mathbf{k})e^{ikx}),
\end{aligned}\end{equation}
where $a^{\mu\dag}$ and $a^{\mu}$ are creation and annihilation operators respectively and they obey the commutation rule

\begin{equation}\label{conmutacion creac aniquilac electromagnetico errado}
[{\widehat{a}}^\mu(\mathbf{k}),{\widehat{a}}^{\beta\dag}(\mathbf{k'})]=\delta^3(\mathbf{k}-\mathbf{k}')\delta^{\mu\beta}.
\end{equation}

With this rule we can write the following commutation relation for the vector fields $A^{\mu}$
\begin{equation}\label{conmutador del acmpo electromagnetico forma covariante}
\left[A^\alpha(x),A^\beta(y)\right]=g^{\alpha\beta}iD_0(x-y),
\end{equation}
where $D_0(z)$ is the Jordan-Pauli distribution (\ref{JPD}) without mass.

For later use in present this work, we compute the vacuum expectation value of the anti-commutator of two $A^{\mu}$ fields
\begin{equation}\label{D1}\begin{aligned}
\langle0|\{A^{\mu}(x),A^{\nu}(y)\}|0\rangle=-g^{\mu\nu}D^{(1)}(y-x),
\end{aligned}\end{equation}
where $D^{(1)}$ is
\begin{equation}\label{D1B}\begin{aligned}
&D^{(1)}(z)=(2\pi)^{-3}\int{d^3}k\delta(k^2)e^{-ikz}.\\
\end{aligned}\end{equation}

We are interested in solving the inhomogeneous differential equation
\begin{equation}\label{EMEL2a}\begin{aligned}
\Box{A}^{\mu}(x)&=g^{\mu}(x).\\
\end{aligned}\end{equation}

Following the Green function method, we get the next retarded and advanced Green functions
\begin{equation}\label{RET}\begin{aligned}
D^{ret}(x-y)
&=(2\pi)^{-4}\int{}d^4pe^{-ip(x-y)}\frac{1}{-p^2-ip^{0}0^{+}},
\end{aligned}\end{equation}

\begin{equation}\label{ADV}\begin{aligned}
D^{adv}(x-y)
&=(2\pi)^{-4}\int{}d^4pe^{-ip(x-y)}\frac{1}{-p^2+ip^{0}0^{+}}.
\end{aligned}\end{equation}
and the solution of (\ref{EMEL2}) is written as
\begin{equation}\label{SolA1}
A=A^{(0)}+\int{}dyD^{ret}(x-y)g^{\mu}(y),
\end{equation}
where
\begin{equation}\label{EMEL2eqret}\begin{aligned}
\Box^{x}D^{ret}(x-y)&=\delta(x-y)\\
\end{aligned}\end{equation}
and $A^{(0)}(x)$ is the solution of the homogeneous differential equation for free electromagnetic fields.

\section{Perturbative method for DKP fields coupled with an Electromagnetic Fields}\label{Kall}

Using gauge invariance, we can see that a Lorentz invariant Lagrangian density for the interaction of DKP fields with electromagnetic fields has the form
\begin{equation}\label{LD}\begin{aligned}
\mathcal{L}
&=\frac{i}{2}(\bar{\psi}\beta^{\mu}\partial_{\mu}\psi-\partial_{\mu}\bar{\psi}\beta^{\mu}\psi)-m\bar{\psi}\psi
-\frac{1}{4}F_{\mu\nu}F^{\mu\nu}-\frac{1}{2}\partial_{\mu}A^{\mu}\partial_{\nu }A^{\nu}+A_{\mu}j^{\mu },
\end{aligned}\end{equation}
where
\begin{equation}\label{TENS}
F^{\mu\nu}=\partial^{\mu}A^{\nu}-\partial^{\nu}A^{\mu}
\end{equation}
and
\begin{equation}\label{CORRP}
j^{\mu }=\frac{e}{2}\{\bar{\psi}_{a}\beta_{ab}^{\mu},\psi_{b}\}.
\end{equation}

The equations of motion for the fields operators in (\ref{LD}) are the coupled differential equations
\begin{equation}\label{EqOpsi}
(i\beta^{\mu}\partial_{\mu}-m)\psi=-e\beta^{\mu}A_{\mu}\psi,
\end{equation}
\begin{equation}\label{EQObarpsi}
\bar{\psi}(i\overleftarrow{\partial}_{\mu}\beta^{\mu}+m)=e\bar{\psi}\beta^{\mu}A_{\mu},
\end{equation}
\begin{equation}
\Box{}A_{\mu}=-\frac{e}{2}\{\bar{\psi}_{a}\beta_{ab}^{\mu},\psi_{b}\}.
\end{equation}

Regarding the solutions (\ref{solgreenge}), (\ref{solgreengebar}) and (\ref{SolA1}), we can write the latter differential equations in their integral form
\begin{equation}\label{solgreenge1}\begin{aligned}
\psi(x)=\psi^{(0)}(x)-\int{d^4y}G_R(x-y)e\beta^{\mu}A_{\mu}(y)\psi(y),
\end{aligned}\end{equation}
\begin{equation}\label{solgreengebar1}\begin{aligned}
\bar{\psi}(x)
&=\bar{\psi}^{(0)}(x)-\int{d^4y}e\bar{\psi}(y)\beta^{\mu}A_{\mu}(y)G_A(y-x),\\
\end{aligned}\end{equation}
\begin{equation}\label{SolA1A}
A=A^{(0)}-\int{}dyD^{ret}(x-y)\frac{e}{2}\{\bar{\psi}_{a}(y)\beta_{ab}^{\mu},\psi_{b}(y)\}.
\end{equation}

To solve the equations (\ref{solgreenge1}), (\ref{solgreengebar1}) and (\ref{SolA1A}) at the operator level is equivalent to solve the problem in the Heisenberg picture. This can be done by expanding the full field operators as a power series in the small coupling constant $e$ as \cite{Kallen1}
\begin{equation}\label{P1}
\psi(x)=\psi^{(0)}(x)+e\psi^{(1)}(x)+e^{2}\psi^{(2)}(x)+\ldots,
\end{equation}
\begin{equation}\label{P3}
\bar{\psi}(x)=\bar{\psi}^{(0)}(x)+e\bar{\psi}^{(1)}(x)+e^{2}\bar{\psi}^{(2)}(x)+\ldots,
\end{equation}
\begin{equation}\label{P2}
A_{\mu}(x)=A_{\mu }^{(0)}(x)+eA_{\mu }^{(1)}(x)+e^{2}A_{\mu }^{(2)}(x)+\ldots.
\end{equation}

Following this methodology, we get the recursion relations below in order to obtain $\psi^{(i)}(x)$, $\bar{\psi}^{(i)}(x)$ and $A^{(i)}_{\mu}(x)$ for $i\geq1$
\begin{equation}\label{RRpsi}\begin{aligned}
\psi^{(n+1)}(x)
&=-\frac{1}{2}\int{d^4y}G_R(x-y)\beta^\mu\sum_{m=0}^{n}\{A_{\mu }^{(m)}(y),\psi^{(n-m)}(y)\},\\
\end{aligned}\end{equation}
\begin{equation}\label{solgreengebarg22}\begin{aligned}
\bar{\psi}^{(n+1)}(x)
&=-\frac{1}{2}\int{d^4y}\sum_{m=0}^{n}\{\bar{\psi}^{(m)}(y),A_{\mu }^{(n-m)}(y)\}\beta^{\mu}G_A(y-x),\\
\end{aligned}\end{equation}
\begin{equation}\label{SolA1gPS1}\begin{aligned}
A_{\mu }^{(n+1)}(x)
&=-\frac{1}{2}\int{}d^{4}yD^{ret}(x-y)\sum_{m=0}^{n}\{\bar{\psi}^{(m)}(y)\beta_{\mu},\psi^{(n-m)}(y)\}.
\end{aligned}\end{equation}
where we have symmetrized in (\ref{RRpsi}) and (\ref{solgreengebarg22}), and $\psi^{(0)}(x)$, $\bar{\psi}^{(0)}(x)$ and $A^{(0)}_{\mu}(x)$ are the free fields operators studied above.

Explicitly, we will write the fields until $i=2$ in terms of free fields for later use
\begin{equation}\label{solgreengeg21}\begin{aligned}
\psi^{(1)}(x)
&=-\frac{1}{2}\int{d^4y}G_R(x-y)\beta^\mu\{A_{\mu }^{(0)}(y),\psi^{(0)}(y)\},\\
\end{aligned}\end{equation}
\begin{equation}\label{solgreengeg22ex}\begin{aligned}
\psi^{(2)}&(x)
=\frac{1}{2}\int{d^4y}\int{d^4z}G_R(x-y)\beta^\mu{G}_R(y-z)\beta^\nu\psi^{(0)}(z)\{A_{\mu }^{(0)}(y),A_{\nu }^{(0)}(z)\}+\\
&+\frac{1}{4}\int{d^4y}\int{}dz^{4}G_R(x-y)\beta^\mu{D}^{ret}(y-z)\times\{\{\bar{\psi}^{(0)}(z)\beta_{\mu},\psi^{(0)}(z)\},\psi^{(0)}(y)\},\\
\end{aligned}\end{equation}
\begin{equation}\label{solgreengebarg21}\begin{aligned}
\bar{\psi}^{(1)}(x)
&=-\frac{1}{2}\int{d^4y}\{\bar{\psi}^{(0)}(y),A_{\mu }^{(0)}(y)\}\beta^{\mu}G_A(y-x),\\
\end{aligned}\end{equation}
\begin{equation}\label{solgreengebarg22ex}\begin{aligned}
\bar{\psi}^{(2)}(x)
&=\frac{1}{4}\int{d^4y}\int{}dzD^{ret}(y-z)\Big\{\bar{\psi}^{(0)}(y),\{\bar{\psi}^{(0)}(z)\beta_{\mu},\psi^{(0)}(z)\}\Big\}\beta^{\mu}G_A(y-x)+\\
&\quad+\frac{1}{2}\int{d^4y}\int{d^4z}\bar{\psi}^{(0)}(z)\{A_{\nu }^{(0)}(z),A_{\mu }^{(0)}(y)\}\beta^{\nu}G_A(z-y)\beta^{\mu}G_A(y-x),\\
\end{aligned}\end{equation}
\begin{equation}\label{SolA1gPS1}\begin{aligned}
A_{\mu }^{(1)}(x)
&=-\frac{1}{2}\int{}dyD^{ret}(x-y)\{\bar{\psi}^{(0)}(y)\beta_{\mu},\psi^{(0)}(y)\},
\end{aligned}\end{equation}
\begin{equation}\label{SolA1gPS12}\begin{aligned}
A_{\mu }^{(2)}(x)
&=\frac{1}{4}\int{}d^{4}y\int{d^4z}D^{ret}(x-y)\big\{\bar{\psi}^{(0)}(y)\beta_{\mu},G_R(y-z)\beta^\mu\{A_{\mu }^{(0)}(z),\psi^{(0)}(z)\}\big\}+\\
&\quad+\frac{1}{4}\int{}d^{4}y\int{d^4z}D^{ret}(x-y)\big\{\{\bar{\psi}^{(0)}(z),A_{\mu }^{(0)}(z)\}\beta^{\mu}G_A(z-y)\beta_{\mu},\psi^{(0)}(y)\big\}.\\
\end{aligned}\end{equation}

Because of experimental interest, we shall compute the expansion of the observable current operator. The expansion of $j^{\mu}$ takes the form
\begin{equation}\label{EXPCurren}
j^{\mu}=j^{(0)\mu}+ej^{(1)\mu}+e^2j^{(2)\mu}+\ldots.
\end{equation}

The terms $j^{(n)\mu}$ are fixed by comparing (\ref{EXPCurren}) with the expression obtained replacing (\ref{P1}) and (\ref{P3}) in (\ref{CORRP})
\begin{equation}\label{4corrienteq0}\begin{aligned}
j^{(0)\mu}(x)&=\frac{e}{2}\{\overline{\psi}^{0}(x)\beta^{\mu},\psi^{0}(x)\},\\
\end{aligned}\end{equation}
\begin{equation}\label{corrente3}\begin{aligned}
j^{(1)\mu}
&=-\frac{e}{2}\int{d^4y}\{\bar{\psi}^{(0)}(x)\beta^{\mu}{G_R}(x-y)\beta^{\nu}A_{\nu}^{(0)}(y),\psi^{(0)}(y)\}-\\
&\quad-\frac{e}{2}\int{d^4y}\{\bar{\psi}^{(0)}(y)\beta^{\nu}A_{\nu}^{(0)}(y){G_A}(y-x)\beta^{\mu},\psi^{(0)}_b(x)\},\\
\end{aligned}\end{equation}
\begin{equation}\label{J2geralC2}\begin{aligned}
j^{(2)\mu}
&=\frac{e}{4}\int{d^4y}\int{d^4z}\{\bar{\psi}^{(0)}(x)\beta^{a},G_R(x-y)\beta^\mu{G}_R(y-z)\beta^\nu\psi^{(0)}(z)\}\{A_{\mu }^{(0)}(y),A_{\nu }^{(0)}(z)\}+\\
&\quad+\frac{e}{8}\int{d^4y}\int{}d^4z\bigg\{\bar{\psi}^{(0)}(x)\beta^{a},G_R(x-y)\beta^\mu{D}^{ret}(y-z)\Big\{j^{(0)}_{\mu}(z),\psi^{(0)}(y)\Big\}\bigg\}+\\
&\quad+\frac{e}{4}\int{d^4y}\int{d^4z}\{\bar{\psi}^{(0)}(z)\beta^{\nu}G_A(z-y)\beta^{\mu}{}G_A(y-x)\beta^{\mu},\psi^{(0)}(x)\}\{A_{\nu }^{(0)}(z),A_{\mu }^{(0)}(y)\}+\\
&\quad+\frac{e}{8}\int{d^4y}\int{}dzD^{ret}(y-z)\bigg\{\Big\{\bar{\psi}^{(0)}(y),j^{(0)}_{\mu}(z)\Big\}\beta^{\mu}G_A(y-x)\beta^{\mu},\psi^{(0)}(x)\bigg\}+\\
&\quad+\frac{e}{8}\int{d^4y}\int{d^4z}\Big\{\{\bar{\psi}^{(0)}(y),A_{\mu }^{(0)}(y)\}\beta^{\mu}{}G_R(x-z)\beta^\mu\{A_{\mu}^{(0)}(z),\psi^{(0)}(z)\}\Big\}.\\
\end{aligned}\end{equation}

\section{Vacuum Polarization}\label{VAC}

To study the vacuum polarization phenomenon, we shall evaluate the vacuum expectation value for the current operator in the presence of an external electromagnetic field $A_{\mu}^{ext}$.

The field $A_{\mu}^{ext}$ is coupled to the DKP particle current modifying the solutions (\ref{solgreenge1}) and (\ref{solgreengebar1}) in the following form
\begin{equation}\label{solgreenge12}\begin{aligned}
\psi(x)&=\psi^{(0)}(x)-\int{d^4y}G_R(x-y)e\beta^{\mu}\Big(A_{\mu}(y)+A_{\mu}^{ext}(y)\Big)\psi(y),
\end{aligned}\end{equation}

\begin{equation}\label{solgreengebar12}\begin{aligned}
\bar{\psi}(x)
&=\bar{\psi}^{(0)}(x)-\int{d^4y}e\bar{\psi}(y)\beta^{\mu}\Big(A_{\mu}(y)+A_{\mu}^{ext}(y)\Big)G_A(y-x).\\
\end{aligned}\end{equation}

{Replacing (\ref{solgreenge12}) and (\ref{solgreengebar12}) in (\ref{4corrienteq0} - \ref{J2geralC2}) and regarding the expansion (\ref{EXPCurren}), we get the current operator $j^{\mu}$ up to first order in the following form:}

\begin{equation}\label{EXPCurrenM}\begin{aligned}
j^{\mu}(x)&=\frac{e}{2}\{\overline{\psi}^{0}(x)\beta^{\mu},\psi^{0}(x)\}\\
&\quad-\frac{e^2}{2}\int{d^4y}\Big\{\bar{\psi}^{(0)}(x)\beta^{\mu}{G_R}(x-y)\beta^{\nu}\big(A_{\nu}^{(0)}(y)+A_{\nu}^{ext}(y)\big),\psi^{(0)}(y)\Big\}-\\
&\quad-\frac{e^2}{2}\int{d^4y}\Big\{\bar{\psi}^{(0)}(y)\beta^{\nu}\Big(A_{\nu}^{(0)}(y)+A_{\nu}^{ext}(y)\Big){G_A}(y-x)\beta^{\mu},\psi^{(0)}(x)\Big\}+\\
&\quad+e^2j^{(2)\mu}+\ldots.
\end{aligned}\end{equation}

{The vacuum expectation value for this current is not null because $A_{\nu}^{ext}$ is not an operator:}
\begin{equation}\label{corrente5}\begin{aligned}
\langle0|j^{\mu}|0\rangle
&=\int{d^4y}K^{\mu\nu}(x-y)A_{\nu}^{ext}(y),\\
\end{aligned}\end{equation}
where $K^{\mu\nu}(x-y)$ is the vacuum polarization tensor which characterizes this phenomenon
\begin{equation}\label{TensK}\begin{aligned}
K^{\mu\nu}(x-y)
&=-\frac{e^2}{2}Tr[G_R(x-y)\beta^{\nu}S^{(1)}(y-x)\beta^{\mu}+S^{(1)}(x-y)\beta^{\nu}G_A(y-x)\beta^{\mu}].\\
\end{aligned}\end{equation}

The interpretation of this non-zero quantity is that the presence of the external field polarizes the vacuum creating a particle and antiparticle represented in the vacuum expectation value $\langle0|\{\bar{\psi}^{(0)}(y),\psi^{(0)}(x)\}|0\rangle$ \cite{Sch} which generates the induced current $j^{(\text{ind})\mu}=\langle0|j^{\mu}|0\rangle$.

The measurable current is given by regarding both the external and the induced current and will be given after the reduction of $K^{\mu\nu}(x-y)$. Regarding the relations (\ref{S1}), (\ref{RetG}), (\ref{AvG}) and using the trace properties for the $\beta$ matrices
\begin{equation}\label{Tr1}
Tr\{\beta^{\mu_1}\beta^{\mu_1}\ldots\beta^{\mu_{2n-1}}\}=0,
\end{equation}
\begin{equation}\label{Tr2}\begin{aligned}
Tr\{\beta^{\mu_1}\beta^{\mu_1}\ldots\beta^{\mu_{2n}}\}
&=g^{\mu_1\mu_2}g^{\mu_3\mu_4}\ldots{}g^{\mu_{2n-1}\mu_{2n}}+g^{\mu_2\mu_3}g^{\mu_4\mu_5}\ldots{}g^{\mu_{2n}\mu_{1}},
\end{aligned}\end{equation}
we get from the Fourier transformation
\begin{equation}\label{TensKFT}\begin{aligned}
\hat{K}^{\mu\nu}(p)
&=(2\pi)^{-2}\int{}d^{4}ze^{ipz}K^{\mu\nu}(z)\\
\end{aligned}\end{equation}
the following reduced expression
\begin{equation}\label{TensKFTb27}\begin{aligned}
\hat{K}^{\mu\nu}(p)
&=-\frac{e^2}{2(2\pi)^{5}m^2}\bigg[-\int{}d^{4}k\int{}d^{4}q\delta(p-k+q)\Big\{[q^2g^{\mu\nu}+q^{\nu}q^{\mu}]\delta(q^2-m^2)+\\
&\quad+[k^2g^{\mu\nu}+k^{\nu}k^{\mu}]\delta(k^2-m^2)\Big\}+\int{}d^{4}k\int{}d^{4}q\Big\{k^2q^{\nu}q^{\mu}+k^{\mu}k^{\nu}q^2+m^2k^{\nu}q^{\mu}+\\
&\quad+m^2k^{\mu}q^{\nu}\Big\}\delta(p-k+q)\Big(\frac{\delta(q^2-m^2)}{k^2-m^2+ik_00^{+}}+\frac{\delta(k^2-m^2)}{q^2-m^2-iq_00^{+}}\Big)\bigg].\\
\end{aligned}\end{equation}

Moreover, because of the conservation of current, the polarization tensor obeys
\begin{equation}\label{Kmunu}
K^{\mu\nu}(p)=[p^{\nu}p^{\mu}-p^2g^{\mu\nu}]B(p^2),
\end{equation}
where
\begin{equation}\label{B1}\begin{aligned}
B(p^2)
&=-\frac{1}{3p^2}K^{\mu}{}_{\mu}(p)\\
&=-\frac{e^2}{3p^22(2\pi)^{5}m^2}\bigg[-5\int{}d^{4}k\int{}d^{4}q\delta(p-k+q)\{q^2\delta(q^2-m^2)+k^2\delta(k^2-m^2)\}+\\
&\quad+2\int{}d^{4}k\int{}d^{4}q\{k^2q^2+m^2k.q\}\delta(p-k+q)\times\\
&\quad\times\Big[\delta(q^2-m^2)\{P\frac{1}{k^2-m^2}-i\pi{}Sgn(k_0)\delta(k^2-m^2)\}+\\
&\quad+\delta(k^2-m^2)\{P\frac{1}{q^2-m^2}+i\pi{}Sgn(q_0)\delta(q^2-m^2)\}\Big]\bigg]\\
\end{aligned}\end{equation}
and $P$ means the principal value and $Sgn$ the sign function. Replacing (\ref{Kmunu}) in (\ref{corrente5}), we get
\begin{equation}\label{corrente51}\begin{aligned}
\langle0|j^{\mu}|0\rangle
&=(2\pi)^{-2}\int{d^4y}\int{d}^4pe^{-i(x-y)p}B(p^2)[\partial^{\mu}\partial^{\nu}A_{\nu}^{ext}(y)-\partial_{\alpha}\partial^{\alpha}A^{ext\mu}(y)]\\
&=-(2\pi)^{-2}\int{d^4y}\int{d}^4pe^{-i(x-y)p}B(p^2){j^{ext}}^{\mu}.\\
\end{aligned}\end{equation}

Now we can compute the observable experimental current ${j^{obs}}^\mu(x)$
\begin{equation}\label{jobs0}\begin{aligned}
{j^{obs}}^\mu(x)
&={j^{ext}}^{\mu}(x)+\langle0|j^{\mu}(x)|0\rangle\\
&=(2\pi)^{-2}\int{d}^4pe^{-ixp}\big[1-(2\pi)^{2}B(p^2)\big]\hat{j}^{ext\mu}(p).\\
\end{aligned}\end{equation}

At this point we can note that if we add a constant $\lambda$ to $(2\pi)^{2}B(p^2)$ and multiply both sides of the equation (\ref{jobs0}) by $(1+\lambda)^{-1}$ , the same physical equation is obtained. Thus, this constant needs to be fixed in order to obtain physical results. Following Faraday's law, we can use the experimental fact that for a slow variation of the external current the induced current is small enough to consider ${j^{obs}}^\mu(x)={j^{ext}}^\mu(x)$. A slow variation of the external current is characterized by the value $p=0$ in (\ref{jobs0}), then ${j^{obs}}^\mu(x)$ is fixed by
\begin{equation}\label{jobs1}\begin{aligned}
{j^{obs}}^\mu(x)
&=(2\pi)^{-2}\int{d}^4pe^{-ixp}\big[1-(2\pi)^{2}B(p^2)+(2\pi)^{2}B(0)\big]\hat{j}^{ext\mu}(p).\\
\end{aligned}\end{equation}

Now we show that this procedure is equivalent to charge renormalization since we are just fixing the units of charge used to define the currents in order to obtain a finite and physical quantity.

With the addition of $(2\pi)^{2}B(0)$ in (\ref{jobs1}), the expression of $K^{\mu\nu}$ is rewritten as
\begin{equation}\label{Kmunu2}
K^{\mu\nu}(p)=[p^{\nu}p^{\mu}-p^2g^{\mu\nu}](B(p^2)-B(0)).
\end{equation}

The quantity $B(p^2)$ can be computed firstly in its imaginary part. From (\ref{B1}) we have
\begin{equation}\label{TFG24}\begin{aligned}
&ImB(p^2)=\\
&=-\frac{e^2}{3p^2(2\pi)^{5}m^2}\bigg[\int{}d^{4}k\int{}d^{4}q\{k^2q^2+m^2k.q\}\delta(p-k+q)\times\\
&\times\Big[-\delta(q^2-m^2)\pi{}Sgn(k_0)\delta(k^2-m^2)+\delta(k^2-m^2)\pi{}Sgn(q_0)\delta(q^2-m^2)\Big]\bigg].\\
\end{aligned}\end{equation}

Regarding Dirac delta function and Lorentz invariance, we can reduce the expression (\ref{TFG24}) to
\begin{equation}\label{TFG29}\begin{aligned}
ImB(p^2)
&=-\pi{}Sgn(p^0)\Pi(p^2),\\
\end{aligned}\end{equation}
where
\begin{equation}\label{PIso}
\Pi(p^2)=\frac{e^2}{192\pi^4}(1-\frac{4m^2}{{p}^2})^{\frac{3}{2}}.
\end{equation}

The real part of $B(p^2)$ can be calculated using the following dispersion relation
\begin{equation}\label{Realpartab}
Ref(\mathbf{p},p^0)=\frac{1}{\pi}P\int\limits_{-\infty}^{\infty}dt\frac{Imf(\mathbf{p},t)}{t-p^0},
\end{equation}

so that
\begin{equation}\label{Realpartab2}\begin{aligned}
ReB(\mathbf{p},p^0)
&=\frac{1}{\pi}P\int\limits_{-\infty}^{\infty}dt\frac{ImB(\mathbf{p},t)}{t-p^0}\\
&=P\int\limits_{0}^{\infty}ds\frac{\Pi(s)}{s-{p}^2}\equiv\bar{\Pi}(p^2).\\
\end{aligned}\end{equation}

Finally, we have
\begin{equation}\label{Bgeral}
B(p^2)=\bar{\Pi}(p^2)-i\pi{}Sgn(p^0)\Pi(p^2)
\end{equation}

From (\ref{PIso}), we can see that for large values of $s$ in (\ref{Realpartab2}), the integral $\bar{\Pi}(p^2)$ is logarithmically divergent, consequently the vacuum polarization tensor in (\ref{Kmunu}) is UV divergent. The addition of $B(0)$, in order to normalize the electric charge, makes $K^{\mu\nu}$ finite

\begin{equation}\label{Kmunua2}\begin{aligned}
K^{\mu\nu}
&=\frac{1}{(2\pi)^4}[p^{\mu}p^{\nu}-p^2g^{\mu\nu}]\frac{e^2p^2}{12}\int\limits_{4m^2}^{\infty}ds\frac{[(1-\frac{4m^2}{s})^{\frac{3}{2}}]}{s(s-{p}^2)}.\\
\end{aligned}\end{equation}

Finally, this last expression shows us the equivalence between Heisenberg and Interaction pictures \cite{Akhiezer} regarding vacuum polarization.

\section{Conclusions and Perspectives}

We have shown that Heisenberg picture and DKP gauge theory lead to the same result for the vacuum polarization tensor that was obtained using Interaction picture. Furthermore, we also showed that the charge renormalization procedure is associated to an experimental fixation for the units of the charge that is used to define the currents related to vacuum polarization.

For future works we hope to study SDKP systematically analyzing its renormalizability and other radiative corrections. We also intend to investigate its total equivalence with SQED in the Heisenberg picture.

%
%

\section*{acknowledgements}
	J. Beltran and N. T. Maia thank to CAPES for full support. B. M. Pimentel thanks to CNPq for partial support.


\begin{thebibliography}{}
	%
	%
	\bibitem{Nakanishi}
	N.~Nakanishi,
	Prog.\ Theor.\ Phys.\,  {\bf 111}, 301 (2004).
	
	\bibitem{Kallen1}
	G. K\"{a}ll\'{e}n. Arkiv f\"{o}r Fysik, {\bf 2}, 187 (1950).
	
	\bibitem{Kallen2}
	G. K\"{a}ll\'{e}n. Arkiv f\"{o}r Fysik {\bf 2} (1950) 371.
	
	\bibitem{Lunardi}
	J.~T.~Lunardi, L.~A.~Manzoni, B.~M.~Pimentel,
	Int.\ J.\ Mod.\ Phys.\ A, {\bf 15}, 3263 (2000).
	
	\bibitem{Lunardi2}
	B.~M.~Pimentel, A.~T.~Suzuki, J.~L.~Tomazelli,
	Int.\ J.\ Theor.\ Phys.\, {\bf 33}, 2199 (1994).
	
	\bibitem{Petiau}
	G. Petiau, Acad. Roy. de Belg. \textbf{16} (1936).
	
	\bibitem{Geheniau}
	J. G\'{e}h\'{e}niau, Acad. Roy. de Belg. \textbf{18} (1938).
	
	\bibitem{Kemmer0}
	N.~Kemmer,
	Proc.\ Roy.\ Soc.\ Lond.\ A {\bf 166} (1938) 127.
	
	\bibitem{Duffin}
	R.~J.~Duffin,
	Phys.\ Rev.\, {\bf 54}, 1114 (1938).
	
	\bibitem{Kemmer}
	N.~Kemmer,
	Proc.\ Roy.\ Soc.\ Lond.\ A, {\bf 173}, 91 (1939).
	
	\bibitem{Nieto}
	R.~A.~Krajcik, M.~M.~Nieto,
	Am.\ J.\ Phys.\, {\bf 45}, 818 (1977).
	
	\bibitem{Tati}
	T. R. Cardoso, B. M. Pimentel, Rev. Bras. Ens. Fis., {\bf 38}, 3 (2016).
	
	\bibitem{CassElDKP}
	J. T. Lunardi, B. M. Pimentel, R. G. Teixeira, J. S.
	Valverde, Phys. Lett.
	A, {\bf 268}, 165 (2000); M. Nowakowski, Phys. Lett.
	A, {\bf 244}, 329 (1998).
	
	
	\bibitem{Okubo}
	S. Okubo, Y. Tosa, Phys. Rev. D, {\bf 20}, 462 (1979).
	
	\bibitem {ClasGrav DKP}
	J.~T.~Lunardi, B.~M.~Pimentel and R.~G.~Teixeira,
	arXiv:gr-qc/9909033 (1999);
	J. T. Lunardi, B. M. Pimentel, R. G. Teixeira, (2001). in Geometrical Aspects of Quantum Fields, proceedings of the 2000 LondrinaWorkshop, Londrina, Brazil, edited by A. A. Bytsenko, A. E. Gonc¸alves and B. M. Pimentel; World Scientific, Singapore p. 111.
	
	\bibitem {FormalDKP-KG}V. Ya. Fainberg, B. M. Pimentel, Theor. Math. Phys.,
	\textbf{124}, 445 (2000); Phys. Lett. A, {\bf 271}, 16 (2000).
	
	\bibitem{Fisch}
	E. Fischbach, M. N. Nieto, C. K. Scott, J. Math. Phys. {\bf 14}, 1760 (1973).
	
	\bibitem{Clark}
	B. C. Clark, S. Hama, G. R. K\"{a}lbermann, R. L. Mercer, L. Ray, Phys. Rev. Lett. {\bf 55}, 592 (1985).
	
	\bibitem{r1}
	R. Casana, J. T. Lunardi, B. M. Pimentel, R. G. Teixeira, General Relativity and Gravitation, {\bf 34}, 1941 (2002).
	
	\bibitem{j1}
	J. T. Lunardi, B. M. Pimentel, R. G. Teixeira, General Relativity and Gravitation, \textbf{34}, 491 (2002).
	
	\bibitem{r2}
	R. Casana, V. Ya. Fainberg, J. T. Lunardi,
	B. M. Pimentel, R. G. Teixeira, Class. Quant. Grav., \textbf{20}, 2457 (2002).
	
	\bibitem{r3}
	R. Casana, J. T. Lunardi, B. M. Pimentel, R. G. Teixeira, Class. Quant. Grav., \textbf{22}, 14 (2005).
	
	\bibitem{a1}
	A. A. Bogush, V. V. Kisel, N. G. Tokarevskaya, V. M. Red’kov, Annales Fond. Broglie, \textbf{32}, 355 (2007).
	
	\bibitem{Umezawa}
	H. Umezawa, Quantum Field Theory (North-Holland Publishing Co., Amsterdam, 1956)
	
	\bibitem{Akhiezer}
	A. I. Akhiezer, V. B. Berestetski, Quantum Electrodynamics, 2nd ed. (Interscience Publisher, New York, 1965)
	
	\bibitem{Kinoshita1}
	T. Kinoshita, Prog. Theor. Phys., \textbf{5}, 473 (1950).
	
	\bibitem{Kinoshita2}
	T. Kinoshita, Y. Nambu, ibid., \textbf{5}, 749 (1950).
	
	\bibitem{Tomazelli}
	B. M. Pimentel, J. L. Tomazelli, Prog. Theor. Phys., \textbf{45}, 1105 (1995).
	
	\bibitem{Gribov}
	V. Gribov, Eur. Phys. J. C, \textbf{10}, 71 (1999).
	
	\bibitem{Kana}
	I. V. Kanatchikov, Rept. Math. Phys. {\bf 46} (2000) 107.
	
	\bibitem{Red}
	V. M. Red’kov, arXiv:quant-ph/9812007 (1998).
	
	\bibitem{Sch}
	J.~S.~Schwinger,
	Phys.\ Rev.\, {\bf 75}, 651 (1948).
\end{thebibliography}


\end{document}